\newcommand{\sg}{\sigma}
\newcommand{\ad}{\mbox{$a^{\diamond}$}}
\newcommand{\bd}{\mbox{$b^{\diamond}$}}
\newcommand{\sgd}{\mbox{$\sigma^{\diamond}$}}
\newcommand{\bs}{\mbox{$b^{{}^{\preceq}}$}}
\newcommand{\fm}{f_{\rm m}}
\newcommand{\pil}{\mbox{$\Pi_{\rm L}$}}
\newcommand{\pir}{\mbox{$\Pi_{\rm R}$}}
\newcommand{\nm}{\!_{\scriptscriptstyle\rm \#}}
\newcommand{\ca}{\mbox{$\mathcal A$}}
\newcommand{\aelem}{\alpha(v)}
\newcommand{\ci}{{\mathcal I}}
\newcommand{\flor}[1]{\lfloor#1\rfloor}
\newcommand{\ceil}[1]{\lceil#1\rceil}
\newcommand{\set}[1]{\{#1\}}
\newcommand{\s}[1]{_{\scriptscriptstyle{\rm #1}}}
\newcommand{\st}[1]{{\scriptscriptstyle #1}}
\newcommand{\thres}[1]{\tau_{#1}}
\newcommand{\addsubspace}{\rule[-1.9ex]{0em}{1ex}}

\documentclass[11pt]{article}

\makeatletter
\def\@begintheorem#1#2{\trivlist
 \item[\hskip\labelsep{\bf #1\ #2.}]\it}
\def\@opargbegintheorem#1#2#3{\trivlist
 \item[\hskip\labelsep{\bf #1\ #2\hskip\labelsep #3.}]\it}

\def\thebibliography#1{\section*{\refname
\@mkboth{\uppercase{\refname}}{\uppercase{\refname}}}%
\small\list{\@biblabel{\arabic{enumiv}}}{%
\setlength{\itemindent}{0em}%
\settowidth\labelwidth{\@biblabel{#1}}%
\leftmargin\labelwidth\advance\leftmargin\labelsep
\setlength{\parsep}{0.9ex plus 0.1ex minus 0.05ex}%
\setlength{\itemsep}{0ex}%
\usecounter{enumiv}%
\let\p@enumiv\@empty%
\def\theenumiv{\arabic{enumiv}}}%
\def\newblock{\hskip .11em plus.11em minus.07em}%
\sloppy\clubpenalty4000\widowpenalty4000}
\def\endthebibliography{%
\def\@noitemerr{\@warning{Empty `thebibliography' environment}}%
\endlist}
\makeatother

\newcommand{\bsection}[2]{%
\section[#1]{\hspace{-1em}.\hspace{0.5em}%
#1}\label{#2}\ignorespaces}

\newtheorem{theorem}{Theorem}
\newcommand{\btheorem}[1]{\begin{theorem}\label{#1}\ignorespaces}
\newcommand{\ethm}{\end{theorem}}

\newtheorem{lemma}[theorem]{Lemma}
\newcommand{\blemma}[1]{\begin{lemma}\label{#1}\ignorespaces}
\newcommand{\elmm}{\end{lemma}}

\newtheorem{corollary}[theorem]{Corollary}
\newcommand{\bcorollary}[1]%
{\begin{corollary}\label{#1}\ignorespaces}
\newcommand{\ecrr}{\end{corollary}}

\newcommand{\bequation}[1]{\unskip\begin{equation}\label{#1}}
\newcommand{\eqtn}{\end{equation}}

\newcommand{\bequations}[1]%
{\unskip\begin{equation}\label{#1}\begin{array}{rcl}}
\newcommand{\eqts}{\end{array}\end{equation}}

\newcommand{\bparagraph}[2]{\par\bigbreak\noindent
\refstepcounter{subsection}\label{#2}{\bf\thesubsection.%
\hspace{0.45em}#1.}\hspace{0.5em}\ignorespaces}

\newcommand{\scnlabel}[1]%
{\noindent{\bf #1.}\hspace{0.5em}\ignorespaces}

\newcommand{\bdisplay}{\begin{eqnarray*}}
\newcommand{\edsl}{\end{eqnarray*}}

\newenvironment{enumroman}{
\begin{list}{(\roman{enumi})}{%
\usecounter{enumi}%
\setlength{\leftmargin}{2\parindent}%
\setlength{\rightmargin}{0em}%
\setlength{\labelsep}{0.75em}%
\setlength{\labelwidth}{\leftmargin}
\addtolength{\labelwidth}{-\labelsep}
\setlength{\listparindent}{\parindent}%
\setlength{\itemindent}{0em}%
\setlength{\partopsep}{0.5\baselineskip plus 0.3ex minus 0.15ex}%
\setlength{\topsep}{0.3ex}%
\setlength{\itemsep}{0ex}%
\setlength{\parsep}{0.3ex}%
}}{\end{list}}

\newcommand{\nitem}[1]{\item\label{#1}\ignorespaces}

\begin{document}%
\title{\mbox{}\\[-1.5cm]%
An In-Place Sorting with\\
  $O(n\log n)$~Comparisons and $O(n)$~Moves%
\footnotetext[1]{Partially supported by the Italian MIUR project
  PRIN ``ALINWEB:\ Algorithmics for Internet and the Web.''}%
\footnotetext[2]{Supported by the Slovak Grant Agen\-cy for
  Science (VEGA) under contract ``Combinatorial Structures and
  Complexity of Algorithms.''}%
}%
\author{Gianni Franceschini\footnotemark[1]\and
  Viliam Geffert\footnotemark[2]\and
\mbox{}\\
{\normalsize \mbox{\footnotemark[1] }\,Department of Informatics
  -- University of Pisa}\\
{\normalsize via Buonarroti 2 -- 56127 Pisa -- Italy}\\
{\normalsize\it francesc@di.unipi.it}\\
{\normalsize \mbox{\footnotemark[2] }\,Department of Computer
  Science -- P.\,J.~\v{S}af\'{a}rik University}\\
{\normalsize Jesenn\'{a} 5 -- 04154 Ko\v{s}ice -- Slovakia}\\
{\normalsize\it geffert@kosice.upjs.sk}%
}%
\date{}%
\maketitle\thispagestyle{empty}%
\begin{quotation}\small
\scnlabel{Abstract}
We present the first in-place algorithm for sorting an array of
size~$n$ that performs, in the worst case, at most $O(n\log n)$
element comparisons and $O(n)$ element transports.

This solves a long-standing open problem, stated explicitly,
e.g., in [J.\,I.~Mun\-ro and V.~Ram\-an, Sorting with minimum data
movement, J.~Algorithms, 13, 374--93, 1992], of whether there
exists a sorting algorithm that matches the asymptotic lower
bounds on all computational resources simultaneously.\\
\mbox{}\\
\scnlabel{Keywords}
Sorting in-place.
\end{quotation}

\bsection{Introduction}{s:i}
{}From the very beginnings of computer science, sorting is one of
the most fundamental problems, of great practical and theoretical
importance. Virtually in every field of computer science there are
problems that have the sorting of a set of objects as a primary
step toward solution. (For early history of sorting,
see~\cite[Sect.~5.5]{Kn73})\@. It is well-known that
a comparison-based algorithm must perform, in the worst case, at
least
$\ceil{\,\log n!\,}\geq n\!\cdot\!\log n -
n\!\cdot\!\log e\approx n\!\cdot\!\log n - 1.443n$
comparisons to sort an array consisting of $n$~elements. (All
logarithms throughout this paper are to the base~$2$, unless
otherwise stated explicitly). By~\cite{MR96}, the corresponding
lower bound for element moves is $\flor{3/2\!\cdot\!n}$\@.

Concerning upper bounds for the number of comparisons, already the
plain version of mergesort gets closely to the optimum, with at
most $n\!\cdot\!\ceil{\log n} - n + 1$ comparisons. However, this
algorithm needs also an auxiliary array for storing $n$~elements,
it is not an {\em in-place} algorithm. That is, it does not work
with only a constant auxiliary storage, besides the data stored in
the input array. In-place algorithms play an important role,
because they maximize the size of data that can be processed in
the main memory without an access, during the computation, to
a secondary storage device.

\medbreak
The rich history of {\em comparisons-storage} family of sorting
algorithms, using $O(n\!\cdot\!\log n)$ comparisons and, at the
same time, $O(1)$ auxiliary storage, begins with a binary-search
version of insertsort. This algorithm uses less than $\log n!+n$
comparisons, only a single storage location for putting elements
aside, and only $O(1)$ index variables, of $\log n$ bits each, for
pointing to the input array. Unfortunately, the algorithm performs
$\Omega(n^2)$ element moves, which makes it unacceptably slow, as
$n$ increases.

The heapsort~\cite{Fl64,Wi64} was the first in-place sorting
algorithm with a total running time bounded by
$O(n\!\cdot\!\log n)$ in the worst case. More precisely, it uses
less than $2n\!\cdot\!\log n$ com\-par\-i\-sons with the same
$O(1)$ storage requirements as insertsort, but only
$n\!\cdot\!\log n+O(n)$ moves, if the moves are organized a little
bit carefully. Since then, many cloned versions of heapsort have
been developed; the two most important ones are
bottom-up-heapsort~\cite{We93} and
a $\log^{\ast}\!$-variant~\cite{Ca92}\@. Both these variants use
not only the same number of moves as the standard heapsort, but
even {\em exactly the same} sequence of element moves for each
input. (See also the procedure ``shiftdown'' in~\cite{SS93})\@.
However, they differ in the number of comparisons. Though
bottom-up variant uses only $3/2\!\cdot\!n\!\cdot\!\log n+O(n)$
comparisons, its upper bound for the average case is even more
important; with $n\!\cdot\!\log n+O(n)$ comparisons, it is one of
the most efficient in-place sorting algorithms. The
$\log^{\ast}\!$-variant is slightly less efficient in an average,
but it guarantees less than
$n\!\cdot\!\log n+n\!\cdot\!\log^{\ast}\!n$ comparisons in the
worst case. For a more detailed analysis, see
also~\cite{LV93,SS93}\@.

Then in-place variants of a $k$-way mergesort came to the
scene~\cite{KPT96,Re92}, with at most $n\!\cdot\!\log n+O(n)$
comparisons, $O(1)$ auxiliary storage, and
$\varepsilon\!\cdot\!n\!\cdot\!\log n+O(n)$ moves. Instead of
merging only $2$ blocks, $k$~sorted blocks are merged together at
the same time. Here $k$ denotes an arbitrarily large, but fixed,
integer constant, and $\varepsilon\!>\!0$ an arbitrarily small,
but fixed, real constant. Except for the first extracted element
in each $k$-tuple of blocks, the smallest element is found with
$\log k$ comparisons, if $k$~is a power of two, since the $k$
currently leftmost elements of the respective blocks are organized
into a selection tree. Though $\log k$ is more than one comparison
required in the standard $2$-way merging, the number of merging
sweeps across the array comes down to $\ceil{\log n/\!\log k}$, so
the number of comparisons is almost unchanged. As an additional
bonus, the number of element moves is reduced if, instead of
elements, only pointers to elements are swapped in the selection
tree. By the use of some other tricks, the algorithm is made
in-place and the size of auxiliary storage is reduced
to~$O(1)$\@. The early implementation of this algorithm, having so
promising upper bounds, turned out to be unacceptably slow. It was
observed that operations with indices representing the current
state of the selection tree became a bottleneck of the program.
Fortunately, the state of a selection tree with a constant number
of leaves can be represented implicitly, without swapping indices.
This indicates that even by summing comparisons and moves we do
not get the whole truth, the arithmetic operations with indices
are also important.

The $k$-way variant has been generalized to a
$(\log n/\!\log\log n)$-way in-place mergesort~\cite{KP99}\@.
This algorithm uses $n\!\cdot\!\log n+O(n\!\cdot\!\log\log n)$
comparisons, $O(1)$ auxiliary storage, and only
$O(n\!\cdot\!\log n/\!\log\log n)$ element moves. Since $k$ is no
longer a constant here, the information about the selection tree
is compressed, among others, into bits of $(\log n)$-bit index
variables by complicated bitwise operations, which increases,
among others, the number of arithmetic operations. Therefore, the
algorithm is mainly of theoretical interest; it is the first
member of the comparisons-storage family breaking the bound
$\Omega(n\!\cdot\!\log n)$ for the number of moves.

\medbreak
The {\em transports-storage} family of algorithms, sorting with
$O(n)$ element moves and $O(1)$ auxiliary storage, is not so
numerous. The first algorithm of this type is selectsort, which is
a natural counterpart of insertsort. Carefully implemented, it
sorts with at most $2n\!-\!1$ moves, a single location for putting
one element aside, and $O(1)$ index variables. Unfortunately, it
performs also $\Omega(n^2)$ comparisons.

As shown in~\cite{MR96}, $O(n^2)$ comparisons and $O(1)$ indices
suffice for reduction of the number of moves to the lower bound
$\flor{3/2\!\cdot\!n}$\@.

Another improvement is a generalized heapsort~\cite{MR92}\@: It
is based on a heap in which internal nodes have $\flor{n^{1/k}}$
children, for a fixed integer~$k$\@. The corresponding heap tree
is thus of constant height, which results in an algorithm sorting
with $O(n)$ moves, $O(1)$ storage, and $O(n^{1+\varepsilon})$
comparisons.

\medbreak
Finally, consider the {\em comparisons-transports} family, sorting
with $O(n\!\cdot\!\log n)$ comparisons and $O(n)$ element moves.
The first member is a so-called tablesort~\cite{Kn73,MR92}\@. We
use any algorithm with $O(n\!\cdot\!\log n)$ comparisons but,
instead of elements, we move only indices pointing to the
elements. When each element's final position has been determined,
we transport all elements to their destinations in linear time.
However, this algorithm requires $\Omega(n)$ auxiliary indices.

The storage requirements have been reduced to
$O(n^{\varepsilon})$ by a variant of samplesort~\cite{MR92}\@.
The same result can also be obtained by the in-place variant of
the $k$-way mergesort~\cite{KPT96,KP99}, mentioned above, if
$k\!=\!\ceil{n^{\varepsilon}}$\@. This reduces the number of
merging sweeps down to a constant, which results in
$O(n\!\cdot\!\log n)$ comparisons and $O(n)$ element moves. Such
modification is no longer in-place, it uses $O(n^{\varepsilon})$
auxiliary indices to represent a selection tree. We leave the
details to the reader.

\medbreak
So far, there was no known algorithm sorting, in the worst case,
with $O(n\!\cdot\!\log n)$ comparisons, $O(n)$ moves, $O(1)$
auxiliary storage, and, at the same time, $O(n\!\cdot\!\log n)$
arithmetic operations.

This ultimate goal has only been achieved in the average
case~\cite{MR92}\@. In the worst case, the algorithm uses
$\Omega(n^2)$ comparisons but, for a randomly chosen permutation
of input elements, the probability of this worst case scenario is
negligible.

It was generally conjectured, for many years, that an algorithm
matching simultaneously the asymptotic lower bounds on all above
computational resources does not exist. For example,
in~\cite{Ra91}, it was proved that the algorithm with
$O(n^{1+\varepsilon})$ comparisons using generalized heaps is
optimal among a certain restricted family of in-place sorting
algorithms performing $O(n)$ moves. It was hoped that, by
generalizing {}from a restricted computational model to all
comparison-based algorithms, we could get a higher trade-off among
comparisons, moves, and storage.

\bparagraph{Our result}{p:our}
The result we shall present in this paper contradicts the above
conjectures and closes a long-standing open problem. We shall
exhibit the first sorting algorithm of the type
{\em comparisons-transports-storage}. Our algorithm operates
in-place, with at most $2n\!\cdot\!\log n + o(n\!\cdot\!\log n)$
element com\-par\-i\-sons and $(13\!+\!\varepsilon)\!\cdot\!n$
element moves in the worst case, for each $n\geq 1$\@. Here
$\varepsilon\!>\!0$ denotes an arbitrarily small, but fixed, real
constant. The number of auxiliary arithmetic operations with
indices is bounded by~$O(n\!\cdot\!\log n)$\@. We can slightly
reduce the number of moves, to $(12\!+\!\varepsilon)\!\cdot\!n$,
in a modified version that uses
$6n\!\cdot\!\log n + o(n\!\cdot\!\log n)$ com\-par\-i\-sons\@.

The algorithm was born as a union of the ideas contained in two
independent technical reports, \cite{Ge02,Fr03}\@. We believe
that, besides the theoretical break\-through achieved by its
analysis, the algorithm can also be of practical interest, because
of its simplicity.

\bparagraph{Algorithm in a nutshell}{p:nut}
Using an evenly distributed sample $a_{1},\ldots,a_{f}$ of size
$\Theta(n/(\log n)^4)$, split the elements into segments
$\sg_{0},\sg_{1},\ldots,\sg_{f}$, of length $\Theta((\log n)^4)$
each, so that elements in~$\sg_{k}$ satisfy
$a_{k}\!\leq\!a\!\leq\!a_{k+1}$\@. The sorted array is obtained by
forming $\sg'_{0},a_{1},\sg'_{1},\ldots,a_{f},\sg'_{f}$, where
$\sg'_{k}$ denotes $\sg_{k}$ in sorted order. To sort~$\sg_{k}$,
use a modified heapsort, with internal nodes having
$\Theta((\log n)^{4/5})$ sons, which results in a constant number
of moves per each element extracted {}from the heap.

Since an evenly distributed sample is hard to find, it grows
dynamically; when some $\sg_{k}$ becomes too large, halve it into
two segments of equal length, and insert the median in the sample.
To minimize moves required for insertions in the sample, it is
sparsely distributed in a block of size $\Theta(n/(\log n)^3)$,
not losing advantage of a quick binary search. A local density of
elements is eliminated by redistributing the sample more evenly,
which does not happen ``too often.'' To avoid the corresponding
segment movement, only pointers connecting $a_{k}$'s
with~$\sg_{k}$'s are moved, the segments stay motionless in
a separate workspace.

However, we do not have a buffer of size~$3n$, required for the
sample and the segments, nor $P\!\approx\!\Theta(n/(\log n)^2)$
bits, for pointers. The bits are ``created'' at the very beginning
by a modified heapsort, collecting the smallest and the largest
$P$ elements to blocks \pil\ and~\pir, which leaves
a block~$\ca'$ in between. Then the $j$th bit can be encoded by
swapping the $j$th element in~\pil\ with the $j$th element
in~\pir\@.

To ``create'' a buffer for sorting the block $\ca'$ of
length~$n'\!$, select the element \bs\ of rank $\flor{n'/4}$ and
partition $\ca'$ into blocks $A\s{<}$ and~$B\s{\geq}$, using \bs\
as a pivot. Then sort~$A\s{<}$, using $B\s{\geq}$ as an empty
buffer. (We can test if a given location contains a buffer
element, by a single comparison with~\bs$\!$\@. Before an
``active'' element is moved, one buffer element escapes to the
current location of the hole). After sorting~$A\s{<}$ we iterate,
focusing on $B\s{\geq}$ as a new block~$\ca'\!$\@. After
$O(\log n)$ iterations, we are done.

\bsection{Sorting with an Additional Memory}{s:sam}
Before presenting our in-place algorithm, we shall concentrate on
a simpler task. We are going to sort a given contiguous
block~$A$, consisting of $m$~elements, using only
$O(m\!\cdot\!\log m)$ comparisons and $O(m)$ element moves. As
some additional resources, we are given a {\em buffer memory}, of
size at least $3m\!-\!1$, that can be used as a temporary
workspace, and a {\em pointer memory}, capable of containing at
least $\flor{4m/(\log m)^2}$ bits.

To let the elements move, we also have a {\em hole}, that is, one
location, the content of which can be modified without destroying
any element. An assignment $a_j\!:=\!a_i$ transports not only one
element {}from the location $i$ to~$j$, but also the hole {}from
$j$ to~$i$\@. At the very beginning, the hole is in a single extra
location, besides the given input array.

\bparagraph{Buffer memory}{p:buf}
The buffer memory forms a separate contiguous block~$B$, initially
consisting of at least $3m\!-\!1$ {\em buffer elements}. All
buffer elements are greater than or equal to a given {\em buffer
separator}~\bs$\!$, placed in an extra location, while all
elements in~$A$ are strictly smaller than~\bs$\!$\@. During the
computation, the elements of $A$ and~$B$ are mixed up. However, by
a single comparison with~\bs$\!$, we can test whether any given
location contains a buffer element, or an {\em active element},
a subject of sorting, placed originally in~$A$\@.

The buffer memory~$B$ consists of two parts. First, there is a low
level {\em segment memory}, a sequence of {\em segments} allocated
dynamically {}from the right end of~$B$ and growing to the left,
as the computation demands. All allocated segments are of the same
fixed length. Second, there is a fixed high level {\em frame
memory}, placed at the left end of~$B$\@.

\bparagraph{Structure of the segment memory}{p:seg}
All segments are of a fixed length~$s$, where
\bequation{e:s}
  s = \left\{
    \begin{array}{l}
      \ \ \ceil{(\log m)^4}\addsubspace\\
      \ \ \ceil{(\log m)^4} + 1
    \end{array}
  \right.
  \mbox{\ \ \ so that $s$ is odd}.
\eqtn
During the computation, the number of active segments never
exceeds $s\nm$, defined by
\bequation{e:sn}
  s\nm = \flor{2m/s}\leq 2m/(\log m)^4,
\eqtn
and hence the size of workspace reserved for the segment memory is
bounded by
\bequation{e:sb}
  S = s\nm\!\cdot\!s\leq 2m\,.
\eqtn
Here we assume that $m$~is ``sufficiently large,'' such that
$s\!\leq\!m$, and hence $s\nm\!\geq\!2$\@. We shall later discuss
how to handle a block~$A$ that is ``short.''

Initially, all segments are {\em free}, containing buffer elements
only. The algorithm keeps the starting position of the last
segment that has been allocated in a global index
variable~$\vec{s}$\@. Initially, $\vec{s}$~points to the right end
of the buffer memory~$B$\@. To allocate a new segment, the
procedure simply performs the operation $\vec{s}:=\vec{s}\!-\!s$,
and returns the new value of $\vec{s}$ as the starting position of
the new segment. Immediately after allocation, some $\flor{s/2}$
active elements (smaller than~\bs) are transported to the first
$\flor{s/2}$ positions of the new segment. The corresponding
buffer elements are saved in the locations released by the active
elements. {}From this point forward, the segment becomes
{\em active}.

In general, the structure of an active segment is
$c_{1}\ldots c_{h}b_{h+1}\ldots b_{s}$, where $c_{1}\ldots c_{h}$
are active elements stored in the segment, while
$b_{h+1}\ldots b_{s}$ are some buffer elements. The value of~$h$
is kept between $\flor{s/2}$ and $s\!-\!1$, so that at least one
half (roughly) of elements in each active segment is active, and
still there is a room for storing one more active element. Neither
$c_{1}\ldots c_{h}$ nor $b_{h+1}\ldots b_{s}$ are sorted. In
addition, the algorithm does not keep any information about the
boundary~$h$ separating active and buffer elements, if the segment
is not being manipulated at the present moment. However, since all
active elements are strictly smaller than~\bs\ and all buffer
elements are greater than or equal to~\bs$\!$, we can quickly
determine the number of active elements in any given segment,
using a binary search with~\bs\ over the $s$ locations of the
segment, which costs only $1\!+\!\flor{\log s}\leq O(\log\log m)$
comparisons, by~(\ref{e:s})\@.

\bparagraph{Structure of the frame memory}{p:fra}
The frame memory, placed at the left end of~$B$, consists of
$r\nm$ so-called {\em frame blocks}, each of length~$r$, where
\bequation{e:r}\begin{array}{rclll}
  r &=& 1 + \ceil{\log(2m/s)}
    &\leq 2\!+\!\log(2m/8) &= \log m\,,\\
  r\nm &=& 2^{r-1} = 2^{\ceil{\log(2m/s)}}
    &\leq 2\!\cdot\!2m/s &\leq 4m/(\log m)^4,
\end{array}\eqtn
using~(\ref{e:s}) and $m\!\geq\!4$\@. That is, the frame memory
is of total length
\bequation{e:rb}
  R = r\nm\!\cdot\!r\leq 4m/(\log m)^3.
\eqtn

Using~(\ref{e:sb}) and $m\!\geq\!4$, we get that the total space
requirements for the segment and frame memories do not exceed the
size of the buffer~$B$, since
$R\!+\!S\leq 4m/(\log m)^3+2m\leq 3m\!-\!1$\@.

A frame block is either {\em free}, containing buffer elements
only, or it is {\em active}, containing some active elements
followed by some buffer elements. Initially, all frame blocks are
free. During the computation, active frame blocks are concentrated
in a contiguous left part of the frame, followed by some free
frame blocks in the right part. However, there are some important
differences {}from the segment memory structure:

First, the active elements, forming a left part of a frame block,
are in sorted order. So are the active frame blocks, forming
a left part of the frame memory. More precisely, let
$a_{1},a_{2},\ldots,a_{f}$ denote the sequence of all active
elements stored in the frame memory, obtained by reading active
elements {}from left to right, ignoring buffer elements and frame
block boundaries. Then $a_{1},a_{2},\ldots,a_{f}$ is a sorted
sequence of elements. Consequently, a subsequence of these, stored
in the first (leftmost) positions of active frame blocks, denoted
here by $a_{i_{1}},a_{i_{2}},\ldots,a_{i_{g}}$, must also be
sorted. Here $f$~denotes the total number of active elements in
the frame, while $g$~the number of active frame blocks, at the
given moment. Similarly,
$a_{i_{j}}a_{i_{j}+1}a_{i_{j}+2}\ldots a_{i_{j+1}-1}$, the
sequence of active elements stored in the $j$th frame block, is
also sorted.

Second, the number of active elements in an active frame block can
range between $1$ and~$r\!-\!1$\@. That is, we keep room for
potential storing of one more active element in each active frame
block, but we do not care about a sparse distribution of active
elements in the frame. The only restriction follows {}from the
fact that there are no free blocks in between some active blocks.

\bparagraph{Relationship between the frame and segments}{p:rel}
Each active element in the frame memory, i.e., each of the
elements $a_{1},a_{2},\ldots,a_{f}$, has an associated segment
$\sg_{1},\sg_{2},\ldots,\sg_{f}$ in the segment memory. The
segment~$\sg_{k}$, for $k$ ranging between $1$ and~$f$, contains
some active elements satisfying $a_{k}\!\leq\!a\!\leq\!a_{k+1}$,
taken {}from~$A$ and stored in the structure so far. The active
elements satisfying $a_{f}\!\leq\!a$ are stored in~$\sg_{f}$,
similarly, those satisfying $a\!\leq\!a_{1}$ are stored in
a special segment~$\sg_{0}$\@. Note that the segment $\sg_{0}$ has
no ``parent'' in the sequence $a_{1},a_{2},\ldots,a_{f}$, that is,
no frame element to be associated with. Chronologically,
$\sg_{0}$~is the first active segment that has been allocated. If
$f\!=\!0$, i.e., no active elements have been stored in the frame
yet, all active elements are transported {}from~$A$
to~$\sg_{0}$\@.

Note also that (in order to keep the number of active elements in
active segments balanced) we do allow some elements equal
to~$a_{k}$ be stored both in $\sg_{k-1}$ and in~$\sg_{k}$\@. In
general, we may even have $a_{k}=a_{k+1}=\ldots=a_{k'}$, for some
$k\!<\!k'\!$\@. Then elements equal to $a_{k}$ may be found in any
of the segments $\sg_{k-1},\sg_{k},\ldots,\sg_{k'}$\@. However,
the algorithm tries to store each ``new'' active element~$a$,
coming {}from~$A$, in the leftmost segment that can be used at the
moment, i.e., it searches for $k$ satisfying
$a_{k}\!<\!a\!\leq\!a_{k+1}$\@.

\medbreak
Recall that we also maintain the invariant that each active
segment contains at least $\flor{s/2}$ active elements. Thus, if
the frame contains $f$ active elements at the given moment,
namely, $a_{1},a_{2},\ldots,a_{f}$, for some $f\!\geq\!1$, the
total number of active elements, stored both in the frame and the
segments $\sg_{0},\sg_{1},\sg_{2},\ldots,\sg_{f}$, is at least
$f+(f\!+\!1)\cdot\flor{s/2}$\@. Now, using the fact that $s$~is
odd, by~(\ref{e:s}), we get that this number is at least
$f+(f\!+\!1)\cdot(s/2\!-\!1/2)=
(f\!+\!1)\cdot s/2+(f/2\!-\!1/2)\geq (f\!+\!1)\cdot s/2$\@.
However, the total number of all active elements is exactly equal
to~$m$, which gives $m\geq (f\!+\!1)\cdot s/2$, and hence also
$f\!+\!1\leq 2m/s$\@. But~$f\!+\!1$, the number of active
segments, is an integer number, which gives that
$f\!+\!1\leq\flor{2m/s}$\@. Therefore, using (\ref{e:sn})
and~(\ref{e:r}),
\bequations{e:f}
  f\!+\!1 &\leq& \flor{2m/s}= s\nm\,,\\
  f &\leq& \flor{2m/s}\leq 2^{\ceil{\log(2m/s)}}= r\nm\,.
\eqts
(The argument has used the assumption that $f\!\geq\!1$\@.
However, (\ref{e:f})~is trivial for $f\!=\!0$, since
$r\nm\!\geq\!s\nm\!\geq\!2$, if $m$ is sufficiently large).

As a consequence, we get that $f\!+\!1$, the number of active
segments, does not exceed~$s\nm$, the capacity of the segment
memory. Second, $f$, the number of active elements in the frame,
will never exceed~$r\nm$, the total number of blocks in the frame,
and hence there is enough room to store all active frame elements,
even if each active frame block contained only a single element of
the sequence $a_{1},a_{2},\ldots,a_{f}$\@.

\bparagraph{Structure of the pointer memory}{p:poi}
The relative order of active frame elements in the sequence
$a_{1},a_{2},\ldots,a_{f}$ does not correspond to the
chronological order, in which the segments
$\sg_{0},\sg_{1},\sg_{2},\ldots,\sg_{f}$ are allocated in the
segment memory. Therefore, with each element position in the
frame, we associate a {\em pointer} to the starting position of
corresponding segment. More precisely, if the frame is viewed as
a single contiguous zone of elements $x_{1}\ldots x_{R}$ (ignoring
boundaries between the frame blocks), then the corresponding zone
of pointers is $\pi_{1}\ldots\pi_{R}$\@. If, for some~$\ell$, the
element~$x_{\ell}$ is a buffer element, then $\pi_{\ell}\!=\!0$,
which represents a {\em NIL} pointer. Conversely, if $x_{\ell}$ is
an active element belonging to the sequence
$a_{1},a_{2},\ldots,a_{f}$, then the value of~$\pi_{\ell}$
represents the starting position of the segment associated
with~$x_{\ell}$\@. (The pointer~$\pi_{0}$ to the
segment~$\sg_{0}$, having no ``parent'' in the frame, is stored
separately, in a global index variable).

Since there are at most $s\nm$ segments, all of equal length,
a pointer to a segment can be represented by an integer value
ranging between $0$ and $s\nm=\flor{2m/s}\leq m/2$,
using~(\ref{e:sn})\@. Thus, a single pointer can be represented
by a block of $p$~bits, where
\bequation{e:p}
  p = 1 + \flor{\log s\nm}\leq \log m\,.
\eqtn
The number of pointers is clearly equal to~$R$, the total size of
the frame. Therefore,
\[
  p\nm = R\,.
\]
Thus, the pointer memory can be viewed as a contiguous array
consisting of $p\nm$ bit blocks, of $p$ bits each, and hence,
by~(\ref{e:rb}), its total length is at most
\bequation{e:pb}
  P = p\nm\!\cdot\!p = R\!\cdot\!p\leq \flor{4m/(\log m)^2}\,,
\eqtn
using also the fact that $P$ must be an integer number.

Since an in-place algorithm can store only a limited amount of
information in index variables, the pointer memory is actually
simulated by two separate contiguous blocks \pil\ and~\pir, each
containing at least $\flor{4m/(\log m)^2}$ elements. Initially,
\pil\ and~\pir\ are sorted, and the largest (rightmost) element in
\pil\ is strictly smaller than the smallest (leftmost) element
in~\pir\@. This allows us to encode the value of the $j$th bit,
for any $j$ ranging between $1$ and $\flor{4m/(\log m)^2}$, by
swapping the $j$th element of \pil\ with the $j$th element
of~\pir\@. Testing the value of the $j$th bit is thus equivalent
to comparing the relative order of the corresponding elements in
\pil\ and~\pir, which costs only a single comparison. Setting
a single bit value requires a single comparison and, optionally,
a single swap of two elements, i.e., $3$ element moves. The
initial distribution of elements in \pil\ and~\pir\ represents all
$\flor{4m/(\log m)^2}$ bits cleared to zero.

\bparagraph{Inserting elements in the structure}{p:ins}
The procedure sorting the block~$A$ works in two phases. In the
first phase, the procedure takes, one after another, all
$m$~active elements {}from~$A$ and inserts them in the structure
described above. The procedure also saves some buffer elements
{}from~$B$, and keeps the structure ``balanced.'' In the second
phase, all active elements are transported back to~$A$, this time
in sorted order.

\medbreak
For each active element $a$ in~$A$, we find a segment, among
$\sg_{0},\sg_{1},\sg_{2},\ldots,\sg_{f}$, where this element
should go.

First, by the use of a binary search with the given element~$a$
over $a_{i_{2}},\ldots,a_{i_{g}}$, that is, over the leftmost
locations in the active frame blocks, find the ``proper'' frame
block for the element~$a$, i.e., the index~$j$ satisfying
$a_{i_{j}}\!<\!a\!\leq\!a_{i_{j+1}}$\@. Note that the
element~$a_{i_{1}}$ is excluded {}from the range of the binary
search. If $a\!\leq\!a_{i_{2}}$, the binary search will return
$j\!=\!1$, i.e., the first frame block. Similarly, for
$a_{i_{g}}\!<\!a$, the binary search returns $j\!=\!g$, i.e., the
last frame block. If $g\!<\!2$, we can go directly to the first
(and only) active frame block without using any binary search,
that is, $j\!:=\!1$\@.

Second, by the use of a binary search with the given element~$a$
over the $r$~locations in the $j$th active frame block, find the
``proper'' active frame element for the element~$a$, i.e., the
index~$k$ satisfying $a_{k}\!<\!a\!\leq\!a_{k+1}$\@. Note that,
since $a_{i_{j}}\!<\!a\!\leq\!a_{i_{j+1}}$, the elements $a_{k}$
and~$a_{k+1}$ are between $a_{i_{j}}$ and~$a_{i_{j+1}}$ in the
sequence $a_{1},a_{2},\ldots,a_{f}$ of all frame elements, not
excluding the possibility that $a_{i_{j}}\!=\!a_{k}$, and/or
$a_{k+1}\!=\!a_{i_{j+1}}$\@. Recall that the $j$th active frame
block begins with the active elements
$a_{i_{j}}a_{i_{j}+1}a_{i_{j}+2}\ldots a_{i_{j+1}-1}$, followed by
some buffer elements, to fill up the room, so that the length of
the block is exactly equal to~$r$\@. These buffer elements are not
sorted, however, they are all greater than or equal to~\bs$\!$,
the smallest buffer element. On the other hand, the element~$a$,
being active, is strictly smaller than~\bs$\!$\@. This allows us
to use the binary search with the given~$a$ in the standard way,
which returns the index~$k$ satisfying
$a_{k}\!<\!a\!\leq\!a_{k+1}$\@. For $a_{i_{j+1}-1}\!<\!a$, the
binary search returns correctly $k\!=\!i_{j+1}\!-\!1$\@. If
$j\!=\!1$, that is, if we are in the first frame block, the binary
search may end up with $k\!=\!0$, indicating that
$a\!\leq\!a_{1}\!=\!a_{i_{1}}$\@.

Third, let the active frame element~$a_k$, satisfying
$a_{k}\!<\!a\!\leq\!a_{k+1}$, be placed in a position~$\ell$ of
the frame memory, that is, $a_{k}\!=\!x_{\ell}$\@. (For
$k\!=\!0$, we take $\ell\!:=\!0$)\@. Then read the information
{}from~$\pi_{\ell}$ in the pointer memory and compute the starting
position of the segment~$\sg_{k}$\@. This segment contains
elements ranging between $a_{k}$ and~$a_{k+1}$\@. If $k\!=\!0$,
i.e., the element~$a$ should go to~$\sg_{0}$, the starting
position of the segment is obtained {}from a separate global index
variable.

Fourth, by the use of a binary search with the buffer
separator~\bs\ over the $s$~locations in the current segment, find
the boundary~$h$ dividing the segment into two parts, namely,
$c_{1}\ldots c_{h}$, the active elements stored in the segment,
and $b_{h+1}\ldots b_{s}$, some buffer elements, filling up the
room.

Fifth, save the buffer element~$b_{h+1}$ aside, to the current
location of the hole, and, after that, store the given
element~$a$ in the segment. If $h\!+\!1\!<\!s$, we are ready to
insert the next element {}from~$A$\@. However, if
$h\!+\!1\!=\!s$, the current segment cannot absorb any more
elements. Therefore, if the segment has become full, we call
a procedure ``rebalancing'' the structure before trying to store
the next element. This procedure will be described later, in
Sect.~\ref{p:rebs}\@.

The above process is repeated until all $m$ active elements have
been inserted in the structure.

\medbreak
Initially, the procedure allocates the segment~$\sg_{0}$, and
stores the first $s\!-\!1$ active elements directly in~$\sg_{0}$,
without travelling via the frame. The number of moves for these
elements is the same as in the standard case, i.e., two moves per
each inserted element.

Let us now determine the standard cost of inserting a single
element. The binary search looking for a proper frame block
inspects a range consisting of $g\!-\!1\!<\!r\nm$ elements, and
hence it performs at most $1\!+\!\flor{\log r\nm}\leq\log m$
comparisons, by~(\ref{e:r})\@. The second binary search, looking
for a proper active element within the given frame block, inspects
a range of $r$ elements, performing at most
$1\!+\!\flor{\log r}\leq O(\log\log m)$ comparisons,
using~(\ref{e:r})\@. Reading the value encoded in the
pointer~$\pi_{\ell}$ requires $p\!\leq\!\log m$ element
comparisons, by~(\ref{e:p})\@. The binary search with~\bs\ over
the $s$~locations in the current segment uses
$1\!+\!\flor{\log s}\leq O(\log\log m)$ comparisons,
by~(\ref{e:s})\@. Finally, saving one buffer element and
transporting the element~$a$ to the current segment can be
performed with $2$~element moves. However, these costs do not
include rebalancing. Since $m$~elements are inserted this way, we
get:

\blemma{l:ins}
If we exclude the costs of rebalancing, inserting $m$ elements in
the structure requires
$2m\!\cdot\!\log m + O(m\!\cdot\!\log\log m)$ comparisons and
$2m$ moves.
\elmm

\bparagraph{Extracting in sorted order\,---\,frame level}{p:extf}
In the second phase, the active elements are transported back
to~$A$, in sorted order. Let $\fm$ denote the maximal value
of~$f$, corresponding to the number of active elements in the
frame at the moment when the last active element has been stored
in the structure. Thus, the frame memory contains the sorted
sequence of active elements $a_{1},a_{2},\ldots,a_{\fm}$,
intertwined with some buffer elements, so the total size of the
frame is~$R$, consisting of elements $x_{1}\ldots x_{R}$\@. Then
we have active elements in the segments
$\sg_{0},\sg_{1},\sg_{2},\ldots,\sg_{\fm}$, with $\sg_{k}$
containing active elements that satisfy
$a_{k}\!\leq\!a\!\leq\!a_{k+1}$\@. Thus, to produce the sorted
order of all active elements, it is sufficient to move, back
to~$A$, the sequence
$\sg'_{0},a_{1},\sg'_{1},a_{2},\sg'_{2},\ldots
,a_{\fm},\sg'_{\fm}$,
where $\sg'_{k}$ denotes the block of sorted active elements
contained in~$\sg_{k}$\@.

\medbreak
The procedure begins with moving the block~$\sg'_{0}$ to~$A$\@.
(We shall return to the problem of sorting a given
segment~$\sg_{k}$ below, in Sect.~\ref{p:exts})\@.

Then, in a loop iterated for $\ell=1,\ldots,R$, check whether
$x_{\ell}$ is an active element. This requires only a single
comparison, comparing $x_{\ell}$ with~\bs$\!$\@. If $x_{\ell}$~is
a buffer element, it is skipped, we can go to the next element in
the frame.

If $x_{\ell}$~is an active element, i.e., $x_{\ell}\!=\!a_{k}$,
for some~$k$, the procedure saves the leftmost buffer element, not
moved yet {}from the output block~$A$, in the current location of
the hole and, after that, moves $x_{\ell}\!=\!a_{k}$ to~$A$\@.
(The first free position in~$A$, i.e., the position of the
leftmost buffer element, is kept in a separate global index
variable, and incremented each time a new active element is
transported back to~$A$)\@. Then we read the value encoded in the
pointer~$\pi_{\ell}$ and compute the starting position of the
segment~$\sg_{k}$\@. After that, we move all active elements
contained in~$\sg_{k}$ to~$A$, in sorted order, by the procedure
presented in Sect.~\ref{p:exts}\@.

\medbreak
Before showing how the segment~$\sg_{k}$ can be sorted, let us
derive computational costs of the above procedure, not including
the cost of sorting~$\sg_{k}$\@. Testing whether $x_{\ell}$ is an
active element, for $\ell=1,\ldots,R$, requires
$R\leq O(m/(\log m)^3)$ comparisons, by~(\ref{e:rb})\@.
Transporting $x_{\ell}\!=\!a_{k}$ to~$A$ requires only $2\fm$
element moves in total, since only active elements are moved. This
gives $2\fm\leq 2r\nm\leq O(m/(\log m)^4)$ element moves, by
(\ref{e:f}) and~(\ref{e:r})\@. Reading the values of
$\fm$~pointers, of length~$p$ bits each, can be done with
$\fm\!\cdot\!p\leq r\nm\!\cdot\!p\leq O(m/(\log m)^3)$
comparisons, using (\ref{e:f}), (\ref{e:r}),
and~(\ref{e:p})\@. Summing up, we have:

\blemma{l:extf}
If we exclude the costs of sorting the segments, extracting in
sorted order requires $O(m/(\log m)^3)$ comparisons and
$O(m/(\log m)^4)$ moves.
\elmm

\bparagraph{Extracting in sorted order\,---\,segment
level}{p:exts}
Now we can describe the routine extracting, in sorted order, all
active elements contained in the given segment~$\sg_{k}$\@. Let
$h_{k}$ denote the number of active elements in~$\sg_{k}$\@.
Clearly, $h_{k}\leq s\leq\ceil{(\log m)^4}\!+\!1$,
using~(\ref{e:s})\@. Initially, the routine determines the value
of~$h_{k}$ by the use of a binary search with~\bs\ over the $s$
locations of the segment. This costs
$1\!+\!\flor{\log s}\leq O(\log\log m)$ comparisons.

\medbreak
After that, the routine uses a generalized version of heapsort,
which in turn uses a modified heap-like structure, with
\[
  t = \ceil{(\log m)^{4/5}}
\]
root nodes (instead of a single root node), and with internal
nodes having $t$ sons (instead of two sons). More precisely, we
organize $c_{1}\ldots c_{h_{k}}$, the active elements contained in
the segment, into the implicit structure with the following
properties:

First, the father of the node~$c_{e}$ is the node~$c_{e'}$, where
$e'\!=\!\flor{(e\!-\!1)/t}$, provided that $e'\!\geq\!1$\@. If
$e'\!<\!1$, then $c_{e}$~is one of the root nodes. This implies
that the heap has $t$ roots, and that the sons of~$c_{e}$ are the
nodes $c_{t\cdot e+1},c_{t\cdot e+2},\ldots,c_{t\cdot e+t}$\@. If,
for some $e$ and~$d\!<\!t$, we have
$t\!\cdot\!e\!+\!d\!=\!h_{k}$, the corresponding node $c_{e}$ has
only $d$ sons, instead of~$t$\@. A leaf is a node $c_{e}$ without
any sons, that is, with $t\!\cdot\!e\!\geq\!h_{k}$\@.

Before passing further, note that the heap does not have more than
five levels, since, by travelling to a root {}from~$c_{h_{k}}$, we
get
\[\begin{array}{rcll}
  h^{(1)} &=& \flor{(h_{k}\!-\!1)/t}   &< h_{k}/t\,,\\
  h^{(2)} &=& \flor{(h^{(1)}\!-\!1)/t} &< h_{k}/t^2,\\
  h^{(3)} &=& \flor{(h^{(2)}\!-\!1)/t} &< h_{k}/t^3,\\
  h^{(4)} &=& \flor{(h^{(3)}\!-\!1)/t} &< h_{k}/t^4,\\
  h^{(5)} &=& \flor{(h^{(4)}\!-\!1)/t} &\leq h^{(4)}/t\!-\!1/t<
    h_{k}/t^5\!-\!1/t\leq s/t^5\!-\!1/t^5.
\end{array}\]
If we had $1\!\leq\!h^{(5)}$, then $1\!<\!s/t^5\!-\!1/t^5$, and
hence also $t^5\!<\!s\!-\!1$\@. Now, using
$t\!=\!\ceil{(\log m)^{4/5}}$ and
$s\!\leq\!\ceil{(\log m)^4}\!+\!1$, by~(\ref{e:s}), we would
obtain $\ceil{(\log m)^{4/5}}^5\!<\!\ceil{(\log m)^4}$, which is
a contradiction. To see this, note that, for each real $x\!>\!0$,
$\ceil{x^{4/5}}^5\geq x^4\!$\@. But $\ceil{x^{4/5}}^5$ is an
integer number, and hence $\ceil{x^{4/5}}^5\geq\ceil{x^4}$\@.

The second property of our heap is that, if a node contains an
active element, then this element is not greater than any of its
sons. Note that we do not care about sons of a node containing
a buffer element. (Initially, there are no buffer elements in the
heap. However, when some active elements have been extracted,
buffer elements will fill up the holes).

This heap property is established in the standard way: For
$e=\flor{(h_{k}\!-\!1)/t},\ldots,1$, establish this property in
the positions $e,\ldots,h_{k}$\@. This only requires to determine
whether $c_{e}$ is not greater than the smallest of its sons and,
if necessary, swap the smallest son with~$c_{e}$\@. Processing
a single node this way costs $t$ comparisons and $3$ element
moves. After that, the heap property is re-established for the son
just swapped in the same way. This may activate a further walk, up
to some leaf.

Taking into account that there are $h^{(1)}$ nodes with paths of
lengths $1$, $2$, $3$, or~$4$ (starting {}from the given node and
ending in a leaf), $h^{(2)}$~nodes with paths of lengths $2$,
$3$, or~$4$, $h^{(3)}$~nodes with paths of lengths $3$ or~$4$, and
$h^{(4)}$~nodes with paths of length~$4$, we get that building the
heap costs $t\!\cdot\!\sum_{i=1}^4 h^{(i)}< 2h_{k}$ comparisons
and $3\!\cdot\!\sum_{i=1}^4 h^{(i)}< 6h_{k}/(\log m)^{4/5}$ moves.

\medbreak
After building the heap, the routine transports, $h_{k}$~times,
the smallest element {}from the heap to the output block~$A$\@.
Here the moves are organized as follows. First, save the leftmost
buffer element, not moved yet {}from~$A$, in the current location
of the hole. Then find the smallest element, placed in one of the
$t$ roots, and move this element to~$A$\@. After that, find the
smallest element among the $t$ sons of this root, and move this
element to the node corresponding to its father. Iterating this
process at most five times, we end up with a hole in some leaf.
Now, we are done. The hole in the leaf will be filled up by
a buffer element in the future, as a side effect. (Usually, in the
next iteration, extracting the next smallest element {}from the
heap).

Thus, unlike in the standard version of heapsort, the size of the
heap does not shrink but, rather, some new buffer elements are
inserted into the heap structure, filling up the leaf holes. These
buffer elements are then handled by the extracting routine in the
standard way, as ordinary active elements. Since these elements
may travel down, {}from the leaf level closer to the root level,
a node containing a buffer element may have a son containing
a smaller buffer element. This will do no harm, however, since
each buffer element is strictly greater than any active element,
because of the buffer separator~\bs$\!$\@. Thus, no buffer element
can be extracted {}from the heap as the smallest element in the
first $h_{k}$ iterations, when the routine terminates.

Deriving the costs of the above routine is straightforward. The
routine repeats $h_{k}$ iterations, performing each time at most
$5(t\!-\!1)\leq 5(\log m)^{4/5}$ comparisons and $6$~moves, since
the heap has at most five levels. This gives
$h_{k}\!\cdot\!5(\log m)^{4/5}$ comparisons and $h_{k}\!\cdot\!6$
moves.

\medbreak
Now we can sum the costs of sorting the segment~$\sg_{k}$\@.
Determining the value of~$h_{k}$ costs $O(\log\log m)$
comparisons. Building the heap costs at most $2h_{k}$ comparisons
and $6h_{k}/(\log m)^{4/5}$ moves. Extracting active elements in
sorted order costs $h_{k}\!\cdot\!5(\log m)^{4/5}$ comparisons and
$h_{k}\!\cdot\!6$ moves. Summing up, we get
$h_{k}\!\cdot\!O((\log m)^{4/5})$ comparisons and
$h_{k}\!\cdot\!(6/(\log m)^{4/5}\!+\!6)$ moves.

To obtain the total cost of sorting all segments
$\sg_{0},\sg_{1},\sg_{2},\ldots,\sg_{\fm}$, we use the fact that
$\sum_{k=0}^{\fm}h_{k}\leq m$, since the number of active elements
stored in the segments is bounded by the total number of active
elements. Therefore, the sum over all segments results in the
following upper bounds:

\blemma{l:exts}
Sorting all segments does not require more than
$O(m\!\cdot\!(\log m)^{4/5})$ com\-par\-i\-sons or
$6m + O(m/(\log m)^{4/5})$ moves.
\elmm

Alternatively, we could use the heap structure with parameter
$t\!=\!\ceil{\log m}$\@. This results in a heap with four levels,
instead of five (since $\ceil{x}^4\geq\ceil{x^4}$, for each real
$x\!>\!0$)\@. This reduces the leading factor for the number of
moves {}from $6m$ to~$5m$\@. The price we pay is increasing the
number of comparisons, {}from $o(m\!\cdot\!\log m)$ to
$4m\!\cdot\!\log m + O(m)$\@. The detailed argument is very
similar to the proof for $t\!=\!\ceil{(\log m)^{4/5}}$\@.

\bparagraph{Rebalancing at the segment level}{p:rebs}
This procedure is activated by the routine of Sect.~\ref{p:ins},
inserting a new active element in the structure, when, for
some~$k$, the segment~$\sg_{k}$ has become full, having absorbed
$s$ active elements.

At the moment of activation, some global index variable is
pointing to the starting position of~$\sg_{k}$\@. The procedure
also remembers~$\ell$, the position of the associated active
element~$a_{k}\!=\!x_{\ell}$ in the frame memory, as well as~$j$,
the position of the frame block containing the element~$a_{k}$\@.
We shall call this block the current frame block. (If
$\sg_{k}\!=\!\sg_{0}$, i.e., $k\!=\!0$, there is no associated
element in the frame. Then $\ell\!=\!0$, but we still have the
current frame block, namely, $j\!=\!1$)\@. The above indices were
computed when the latest active element was inserted in the
structure.

\medbreak
First, by the use of a binary search with the buffer
separator~\bs\ over the $r$~locations in the current frame block,
find~$\ell'\!$, the position of the leftmost buffer element in
this block. We shall denote this element by~\bd$\!$\@. Recall that
we maintain the invariant that each active frame block has a room
for one more active element, and therefore it does contain at
least one buffer element.

Second, find a median in the segment~$\sg_{k}$, i.e., an
element~\ad\ of rank~$\flor{s/2}\!+\!1$\@. Without loss of
efficiency, the selection procedure will position \ad\ at the end
of~$\sg_{k}$\@.

Third, the median~\ad\ is inserted in the current frame block, one
position to the right of~$a_{k}$\@. The active elements lying in
between $a_{k}$ and~\bd$\!$, that is, occupying locations
$x_{\ell+1}\ldots x_{\ell'-1}$ in the frame memory, are shifted
one position to the right. At the same time, \bd~is saved
{}from~$x_{\ell'}$ to the location released by~\ad\ at the end of
the segment~$\sg_{k}$\@. (As a special case, if $a_{k}$~is the
rightmost active element in the current frame block, only \bd\
and~\ad\ are swapped. The same holds when $\sg_{0}$~is rebalanced
for the first time, with $\ell\!=\!0$ and $\ell'\!=\!1$). Since
\ad~has been picked {}from~$\sg_{k}$, it satisfies
$a_{k}\!\leq\!\ad\!\leq\!a_{k+1}$, and hence the sequence of
active elements stored in the frame memory remains sorted.

Fourth, after shifting the active elements in the locations
$x_{\ell+1}\ldots x_{\ell'-1}$ one position to the right, we have
to shift the corresponding pointers
$\pi_{\ell+1}\ldots\pi_{\ell'-1}$ as well, so the active elements
remain connected with their segments. To move an integer pointer
value {}from $\pi_{e}$ to~$\pi_{e+1}$, we only have to read the
value encoded in~$\pi_{e}$ and, at the same time,
clear~$\pi_{e}$, and then to encode this value in~$\pi_{e+1}$\@.
Such transport of a pointer costs $O(p)$ comparisons and moves.

Fifth, we need to connect a new active element in the frame with
a new segment. This concerns the element~\ad$\!$, now placed
in~$x_{\ell+1}$\@. Thus, we allocate a new segment~$\sgd$ and
encode its starting position in the pointer~$\pi_{\ell+1}$\@.

Sixth, the full segment~$\sg_{k}$ is halved, that is, we place
some $\flor{s/2}$ active elements greater than or equal to~\ad\
into the left part of~\sgd\ and collect the remaining
$\flor{s/2}$ active elements, smaller than or equal to~\ad$\!$, in
the left part of the original segment~$\sg_{k}$\@. Since many
elements may be equal to~\ad$\!$, we distribute such elements both
to $\sg_{k}$ and~\sgd$\!$, so that their active parts are of equal
lengths. This also requires to save $\flor{s/2}$ buffer elements,
placed originally in~\sgd$\!$, to the locations released
in~$\sg_{k}$\@. (We shall give more details below, in
Sect.~\ref{p:hal})\@. The outcome of halving is that the active
elements in~$\sg_{k}$ are split into two segments $\sg_{k}$
and~\sgd$\!$, satisfying $a_{k}\!\leq\!a\!\leq\!\ad$ and
$\ad\!\leq\!a\!\leq\!a_{k+1}$, respectively.

Seventh, if there is still a room for storing one more active
element in the current frame block, the structure has been
rebalanced. We are done, ready to take the next element
{}from~$A$\@. However, if this block has become full, because
of~\ad$\!$, the program control jumps to a routine rebalancing the
frame level, described later, in Sect.~\ref{p:rebf}\@.

\medbreak
Let us now derive the computational costs. The binary search,
determining the position of the leftmost buffer element in the
current frame block, inspects a range of $r$ elements, performing
$1\!+\!\flor{\log r}\leq O(\log\log m)$ comparisons,
by~(\ref{e:r})\@. Finding a median, in a segment of length~$s$,
requires only $O(s)\leq O((\log m)^4)$ comparisons and
$\varepsilon\!\cdot\!s\leq\varepsilon\!\cdot\!(2\!+\!(\log m)^4)$
element moves, where $\varepsilon\!>\!0$ is an arbitrarily small,
but fixed, real constant, by~\cite{GK01} and~(\ref{e:s})\@.
Rearranging the elements \ad$\!$, \bd$\!$, and
$x_{\ell+1}\ldots x_{\ell'-1}$ in their locations can be done with
at most $r\!+\!2\leq O(\log m)$ moves, by~(\ref{e:r})\@. Shifting
the pointers $\pi_{\ell+1}\ldots\pi_{\ell'-1}$ one position to the
right costs $O(r\!\cdot\!p)\leq O((\log m)^2)$ comparisons, by
(\ref{e:r}) and~(\ref{e:p}), together with the same number of
moves. Encoding the starting position of a new segment in the
pointer~$\pi_{\ell+1}$ requires $O(p)\leq O(\log m)$ element
moves, by~(\ref{e:p})\@. Halving the active elements
in~$\sg_{k}$ into two segments $\sg_{k}$ and~\sgd\ requires only
$O(s)\leq O((\log m)^4)$ comparisons and
$3/2\!\cdot\!s\leq 3/2\!\cdot\!(2\!+\!(\log m)^4)$ moves, using
Lem.~\ref{l:hal}, displayed in Sect.~\ref{p:hal} below,
and~(\ref{e:s})\@.

By summing the bounds above, we get that a single activation of
the procedure rebalancing a segment performs $O((\log m)^4)$
comparisons and $(3/2\!+\!\varepsilon)\!\cdot\!(\log m)^4$ moves.
Taking into account that each activation increases the number of
active segments, that we start with one segment, namely,
$\sg_{0}$, and that we end up with $\fm\!+\!1$ segments, we see
that the number of activations is bounded by~$\fm$\@. This value
is bounded by $\fm\leq s\nm\leq 2m/(\log m)^4$, using
(\ref{e:f}) and~(\ref{e:sn})\@. This gives:

\blemma{l:rebs}
The total cost of keeping the segment level balanced is $O(m)$
comparisons and $(3\!+\!\varepsilon)\!\cdot\!m$ moves, where
$\varepsilon\!>\!0$ is an arbitrarily small, but fixed, real
constant.
\elmm

\bparagraph{Halving a segment}{p:hal}
Here we describe a simple procedure for halving, needed in
Sect.~\ref{p:rebs} above. We are given a segment~$\sg_{k}$ of
size~$s$, and a median~\ad$\!$, that is, an element of rank
$\flor{s/2}\!+\!1$, put aside. We want to place some $\flor{s/2}$
active elements greater than or equal to~\ad\ into the left part
of another given segment~\sgd$\!$, of size~$s$ again, and collect
the remaining $\flor{s/2}$ elements smaller than or equal to~\ad\
in the left part of~$\sg_{k}$\@. The first $\flor{s/2}$ buffer
elements of~\sgd\ must be saved.

\medbreak
In the first phase, with $s\!-\!1$ comparisons and no moves, we
count~$c'\!$, the number of elements strictly smaller
than~\ad$\!$, in~$\sg_{k}$\@. This gives us
$c=\flor{s/2}\!-\!c'\!$, the number of elements equal to~\ad\ that
should remain in~$\sg_{k}$\@. This number will be required in the
second phase, when each element~$a$ of~$\sg_{k}$ is compared
with~\ad\ twice, using ``$a\!<\!\ad$\,'' and
``$a\!>\!\ad$\,''$\!$\@. The elements strictly smaller than~\ad\
and the first $c$ elements detected to be equal to~\ad\ will be
considered ``small,'' while the remaining equal elements and those
strictly greater than~\ad\ will be ``large.'' Each time an element
$a\!=\!\ad$ will be detected, the counter~$c$ will be decreased by
one, until it gets to zero. {}From then on, any ``new''
element~$a$ will be considered ``small'' if and only if
$a\!<\!\ad\!$, and ``large'' otherwise.

In the second phase, the configurations of the segments are
$\sg_{k}=A_1UB_1\bd$ and $\sgd=A_2B_2$, where $A_1$ and~$A_2$
denote, respectively, the active elements of~$\sg_{k}$ found to be
``small'' or ``large,'' collected so far, $B_1$~the buffer
elements moved {}from \sgd\ to~$\sg_{k}$, $B_2$~the elements
of~\sgd\ not moved yet, $U$~the elements of~$\sg_{k}$ not examined
yet, and \bd~a~single buffer element, filling up the room.
$A_2$~and $B_1$ are of equal length, not exceeding
$\flor{s/2}$\@. Initially, $\sg_{k}\!=\!U\bd\!$, $\sgd\!=\!B_2$,
with $A_1,A_2,$ and~$B_1$ empty. The procedure also remembers the
current position of the hole. (After the first iteration, the hole
is always in the leftmost location of~$B_1$)\@.

The second phase proceeds in a loop, as follows. Using at most two
comparisons, the rightmost element~$a$ of~$U$ is determined to be
``small'' or ``large.'' If $a$~is large, we save the leftmost
element {}from~$B_2$ in the current location of the hole and fill
up the new hole in~$B_2$ by~$a$\@. Thus, $A_2$ and~$B_1$ have been
extended, while $U$ and~$B_2$ have been reduced. If $a$~is small,
we scan~$U$ {}from left to right until we find the first
element~$a'$ that is large. All elements on the left of~$a'$
become a part of~$A_1$, without being moved. Since $a$~is a small
element placed on the right of the position~$\flor{s/2}$,
$a'$~must be found before we reach the position
$\flor{s/2}\!+\!1$, or else we would have more than $\flor{s/2}$
small elements, which is a contradiction. Now we save the leftmost
element {}from~$B_2$ to the hole, fill up the hole in~$B_2$
by~$a'\!$, and move~$a$ to the place released by~$a'\!$\@. Then
all necessary boundaries are updated.

This is repeated until we have transported exactly $\flor{s/2}$
active large elements {}from $\sg_{k}$ to~\sgd$\!$\@. As
a consequence, the remaining $\flor{s/2}$ active elements
of~$\sg_{k}$, placed on the left of~$B_1$, must be all small,
since the rank of~\ad\ is $\flor{s/2}\!+\!1$ and $s$~is odd,
by~(\ref{e:s})\@.

\medbreak
Clearly, we have used at most $3s$ comparisons in total, and at
most three moves per each large element moved {}from $\sg_{k}$
to~\sgd$\!$\@. This gives:

\blemma{l:hal}
Given a median \ad$\!$, a segment of size~$s$ can be halved with
at most $O(s)$ com\-par\-i\-sons and $3/2\!\cdot\!s$ moves.
\elmm

\bparagraph{Rebalancing at the frame level}{p:rebf}
This routine is activated by the procedure of
Sect.~\ref{p:rebs}, rebalancing a segment, when it finds out
that, for some~$j$, the $j$th frame block has become full, having
absorbed $r$ active elements. As a side effect, the routine may
increase the number of active blocks in the frame. The routine is
based on a new variant of the well-known data structure
(see~\cite{IKR81,Wi82}), used to maintain a set of elements in
sorted order in a contiguous zone of memory.

\medbreak
For the purpose of keeping the frame memory balanced, the frame
consisting of $r\nm$ frame blocks is viewed, implicitly, as
a complete binary tree with $r\nm\!=\!2^{r-1}$ leaves, and hence
of (edge) height~$r\!-\!1$\@. We introduce the following numbering
of levels: $i\!=\!0$~for the leaves, $1$~for their fathers, and so
on, ending by $i\!=\!r\!-\!1$ for the root. Each node of the tree
is associated with a contiguous subarray of the frame blocks, and
with a path leading to this node {}from the root, as follows.

The $j$th leaf, for any $j$ ranging between $1$ and~$2^{r-1}\!$,
is associated with the $j$th frame block, i.e., with a subarray
consisting of $1\!=\!2^{0}$ frame blocks, starting {}from the
block position~$j$\@. The corresponding path {}from the root to
this leaf is represented by the number $\vec{\jmath}=j\!-\!1$\@.
It is easy to see that by reading the binary representation
of~$\vec{\jmath}$ {}from left to right (with leading zeros so that
its length is~$r\!-\!1$) we get the branching sequence along this
path; $0$~is interpreted as branching to the left, while $1$~as
branching to the right.

Given a node~$v$ at a level~$i$, associated with a path
number~$\vec{\jmath}$ and with a subarray of length~$2^{i}$
blocks, starting {}from a block position~$j$, the father~$v'$ of
this node is associated with the path number
$\vec{\jmath}\,'=\flor{\vec{\jmath}/2}$, and with the subarray of
length~$2^{i+1}\!$, starting {}from the block
position~$j'\!=\!j$, if $\vec{\jmath}$~is even ($v$~is a left son
of~$v'$), but {}from $j'\!=\!j\!-\!2^{i}\!$, if $\vec{\jmath}$~is
odd (right son)\@. Thus, the subarray for the father is obtained
by concatenation of the two subarrays for its sons, while its path
number by cutting off the last bit in the path number for any of
its sons.

During the computation, the number of active elements in some
local area of the frame may become too large. The purpose of
rebalancing a subarray, associated with a node~$v$ at
a level~$i$, for $i\!>\!0$, is to eliminate such local densities
and redistribute active elements more evenly. More precisely,
after rebalancing the subarray, the following two conditions will
hold:

\begin{enumroman}
  \nitem{r:thr} The number of active elements, in any frame block
    belonging to the subarray associated with the given node~$v$
    at the level~$i$, will not exceed the threshold
    $\thres{i}=r\!-\!i$\@.
  \nitem{r:cnt} The frame memory will not contain any free blocks
    (without active elements) in between some active blocks.
\end{enumroman}

Note that, if a node~$v$ at a level $i\!>\!0$ is an ancestor of
the $j$th leaf, the condition~(\ref{r:thr}) ensures that the
$j$th frame block is not full any longer. Neither is any other
block within the subarray. Such redistribution of active elements
is possible only if~$\aelem$, the total number of active elements
in the subarray associated with~$v$, is bounded by
$\aelem\leq\thres{i}\!\cdot\!2^{i}\!$\@. We say that the node~$v$
{\em overflows}, if $\aelem>\thres{i}\!\cdot\!2^{i}\!$\@.

The condition~(\ref{r:cnt}) is required only because of the
procedure presented in Sect.~\ref{p:ins}, transporting active
elements {}from the block~$A$ to the structure. Recall that this
procedure uses a binary search over the leftmost locations in the
active frame blocks, and hence these blocks must form a contiguous
zone.

\medbreak
Now we can describe the routine rebalancing the frame.

First, starting {}from the father of the frame block that is full,
climb up and find the lowest ancestor~$v$ that does not overflow,
with $\aelem\leq\thres{i}\!\cdot\!2^{i}\!$\@. The formulas for
$j$ and~$\vec{\jmath}$, presented above, give us a simple tool for
computing the boundaries of the associated subarrays, along the
path climbing towards the root. To compute the value of~$\aelem$,
for the given ancestor~$v$ at the given level~$i$, simply scan all
$2^{i}$ frame blocks forming the associated subarray and sum up
the numbers of active elements in these blocks, using a binary
search with the buffer separator~\bs\ over the $r$~locations in
each block.

Second, move the $\aelem$ active elements in the associated
subarray of~$v$ to the last $\aelem$ locations. That is,
processing all $2^{i}\!\cdot\!r$ locations in the subarray {}from
the right to left, collect all elements smaller than~\bs\ to the
right end. Before moving an active element {}from $x_{e}$
to~$x_{e'}$, for some $e\!<\!e'\!$, the buffer element in the
target position~$x_{e'}$ is saved to the current location of the
hole. Then move the associated pointer in the corresponding
positions of the pointer memory, {}from $\pi_{e}$ to~$\pi_{e'}$,
by reading and clearing the bit value encoded in~$\pi_{e}$ and
encoding this value in~$\pi_{e'}$\@.

Third, redistribute the $\aelem$~active elements back, this time
more evenly in the $2^{i}$~frame blocks of the subarray, moving
also the pointers in the corresponding positions, as follows. Let
$\alpha\s{D}=\flor{\aelem/2^{i}}$ and
$\alpha\s{M}=\aelem\bmod 2^{i}$\@. Then put $\alpha\s{D}\!+\!1$
active elements in each of the first $\alpha\s{M}$ blocks, and
$\alpha\s{D}$~active elements in each of the remaining
$2^{i}\!-\!\alpha\s{M}$ blocks. In each block, the active elements
are concentrated in its left part.

Fourth, as a side effect of redistribution, the size of the active
part in the frame memory may have been increased. This requires to
update the value of~$g$, the number of active frame blocks, kept
in a separate global index variable. Let $g'$~be the block
position of the rightmost frame block in the subarray of~$v$\@.
Then let $g:=\max\set{g,g'}$\@.

\medbreak
It should be pointed out that, for each leaf, the desired
ancestor~$v$ without overflow does exist. Using (\ref{e:f}),
(\ref{e:r}), and $\thres{i}=r\!-\!i$, for the level $i=r\!-\!1$,
that is, for $v$ being the root node, we get
$\aelem= f\leq r\nm= 1\!\cdot\!2^{r-1}=
\thres{r-1}\!\cdot\!2^{r-1}\!$,
and hence at least the root node does not overflow. Therefore, in
the first step, the loop climbing up towards the root must halt
correctly.

Further, the redistribution of active elements, presented in the
third step, is correct, since
$(\alpha\s{D}\!+\!1)\!\cdot\!\alpha\s{M} +
\alpha\s{D}\!\cdot\!(2^{i}\!-\!\alpha\s{M}) = \aelem$\@.
It is easy to see that the redistribution satisfies the
condition~(\ref{r:thr}) above, using the fact that the node~$v$
does not overflow, and hence
$\aelem\leq\thres{i}\!\cdot\!2^{i}\!$\@. There are two cases to
consider: For $\aelem\leq\thres{i}\!\cdot\!2^{i}\!-\!1$, we have
$\alpha\s{D}\!+\!1\leq
\flor{(\thres{i}\!\cdot\!2^{i}\!-\!1)/2^{i}}\!+\!1\leq
(\thres{i}\!-\!1)\!+\!1= \thres{i}$,
since $\thres{i}$~is an integer. If
$\aelem=\thres{i}\!\cdot\!2^{i}\!$, we get $\alpha\s{M}\!=\!0$,
and hence all $2^{i}$ blocks are ``remaining,'' with only
$\alpha\s{D}$~active elements in each. But here
$\alpha\s{D}= \flor{(\thres{i}\!\cdot\!2^{i})/2^{i}}=
\flor{\thres{i}}= \thres{i}$\@.

It is also easy to see that the redistribution satisfies the
condition~(\ref{r:cnt})\@. Since $v$ is the {\em lowest} ancestor
that does not overflow, along the path {}from the full frame block
towards the root, it must have a son that does overflow, with at
least $\thres{i-1}\!\cdot\!2^{i-1}$ active elements in its
subarray. (As a special case, for $i\!=\!1$, we get
$\thres{0}\!\cdot\!2^{0}=(r\!-\!0)\!\cdot\!1=r$ active elements in
the $j$th frame block that is full). The subarray of the son is
a part of the subarray associated with~$v$, and hence
$\aelem\geq\thres{i-1}\!\cdot\!2^{i-1}\geq 2\!\cdot\!2^{i-1}\!$,
using the fact that $i\!-\!1\!\leq\!r\!-\!2$\@. But then
$\alpha\s{D}=\flor{\aelem/2^{i}}\geq 1$\@. This implies that each
frame block in the subarray associated with~$v$ contains at least
one active element after redistribution, and hence the zone of
active frame blocks will remain contiguous.

\medbreak
Consider now the cost of a single activation of the above routine,
rebalancing a subarray for a node~$v$ at a level~$i$\@. Looking
for the lowest ancestor without overflow requires to count the
numbers of active elements in the associated subarrays along
a path climbing up {}from a father of a leaf, for levels
$e=1,\ldots,i$\@. In the $e$th level, $2^{e}$~blocks are examined,
by a binary search over the $r$~locations of the block.
By~(\ref{e:r}), this gives
$\sum_{e=1}^{i} 2^{e}\!\cdot\!(1\!+\!\flor{\log r})\leq
2^{i}\!\cdot\!O(\log\log m)$
comparisons. The cost of the second step, collecting $\aelem$
active elements to the right end, is $2^{i}\!\cdot\!r$ comparisons
(one comparison with~\bs\ for each location in the subarray), plus
$\aelem\!\cdot\!2\!+\!1$ moves (two moves per each collected
element). However, with each collected element, the corresponding
pointer must also be transported, which gives additional
$\aelem\!\cdot\!O(p)$ comparisons and moves. Using
$\aelem\leq\thres{i}\!\cdot\!2^{i}\leq r\!\cdot\!2^{i}\!$,
together with (\ref{e:r}) and~(\ref{e:p}), the cost of the
second step can be bounded by
$2^{i}\!\cdot\!O(r\!\cdot\!p)\leq 2^{i}\!\cdot\!O((\log m)^2)$
comparisons and moves. The same computational resources are
sufficient in the third step, redistributing the same number of
active elements back, but more evenly, together with their
pointers. Again, this gives $\aelem\!\cdot\!O(p)$ comparisons and
moves, which can be bounded by $2^{i}\!\cdot\!O((\log m)^2)$\@.
Finally, the fourth step does not require any element comparisons
or moves, it just updates one index variable, in $O(1)$ time.

Summing up, the cost of a single activation is
$2^{i}\!\cdot\!O((\log m)^2)$ comparisons and moves, for each node
$v$ at the fixed level $i\!>\!0$\@. To get the total cost, we must
take into account how frequently such rebalancing is activated.

When a rebalancing is activated, $v$~must have a son with at least
$\thres{i-1}\!\cdot\!2^{i-1}$ active elements, since $v$~is the
lowest ancestor that does not overflow, along some path climbing
up. Now, trace back the history of computation, to the moment when
the entire subarray associated with~$v$ was a subject of
redistribution for the last time. This way we get a node~$v'\!$,
either an ancestor of~$v$ or $v$~itself, at a level
$i'\!\geq\!i$, with the associated subarray containing the entire
subarray for~$v$\@. After the redistribution for~$v'\!$, both sons
of~$v$ contained at most
$\thres{i'}\!\cdot\!2^{i-1}\leq\thres{i}\!\cdot\!2^{i-1}$ active
elements. Thus, in the meantime, the number of active elements in
one of the sons of~$v$ has been increased by at least
$\thres{i-1}\!\cdot\!2^{i-1}\!-\!\thres{i}\!\cdot\!2^{i-1}=
2^{i-1}\!$\@.
Since other redistributions, taking place between the moments of
rebalancing $v'$ and~$v$, could not ``import'' any active elements
to the subarray of~$v$ {}from any other parts of the frame, the
$2^{i-1}$ additional active elements must have been {\em inserted}
here. (See the procedure of Sect.~\ref{p:rebs}, third step).
Thus, there have to be at least $2^{i-1}$ insertions in the
associated subarray between any two redistributions for~$v$\@.
Note that, for the fixed level~$i$, subarrays associated with
different nodes~$v$ do not overlap. Thus, we can charge the cost
of each activation, for the given node~$v$, to the $2^{i-1}$
insertions preceding this activation in the given subarray,
without charging the same insertion more than once. This gives
$2^{i}\!\cdot\!O((\log m)^2)/2^{i-1}\leq O((\log m)^2)$
comparisons and moves, per a single insertion of an active element
in the frame memory. Since, in the whole computation, there were
only $\fm\leq r\nm\leq O(m/(\log m)^4)$ insertions, by
(\ref{e:f}) and~(\ref{e:r}), we get the cost $O(m/(\log m)^2)$
comparisons and moves, for rebalancing of all nodes at the fixed
level~$i$\@. By summing over all levels, using
$i\leq r\!-\!1\leq\log m$, by~(\ref{e:r}), we get the total cost:

\blemma{l:rebf}
The total cost of keeping the frame memory balanced is
$O(m/\log m)$ comparisons, together with the same number of moves.
\elmm

\bparagraph{Summary}{p:sum}
By summing the bounds presented in
Lems.~\mbox{\ref{l:ins}\,--\,\ref{l:rebs}} and~\ref{l:rebf}
above, we get:

\btheorem{t:sum}
The cost of sorting the given block~$A$ of size~$m$ is
$2m\!\cdot\!\log m + O(m\!\cdot\!(\log m)^{4/5})$ comparisons and
$(11\!+\!\varepsilon)\!\cdot\!m$ moves, where $\varepsilon\!>\!0$
is an arbitrarily small, but fixed, real constant, provided we can
use additional buffer and pointer memories, of respective sizes
$3m\!-\!1$ and $\flor{4m/(\log m)^2}$\@.
\ethm

The algorithm presented above assumes that $m$~is ``sufficiently
large,'' so that~$s$, defined by~(\ref{e:s}), satisfies
$s\!\leq\!m$\@. This presupposition holds for each
$m\!>\!2^{16}\!=\!65536$\@. Shorter blocks are handled in
a different way, by the procedure described later, in
Sect.~\ref{p:sho}\@. The bounds presented by Thm.~\ref{t:sum}
for the number of comparisons and moves will remain valid.

\bsection{In-Place Sorting}{s:ips}
Now we can present an in-place algorithm sorting the given
array~\ca\ consisting of $n$~elements. If $n\!\leq\!2^{16}\!$, the
array is sorted directly, by the procedure of Sect.~\ref{p:sho},
described later. In the general case, for $n\!>\!2^{16}\!$, the
task of the main program is to provide sufficiently large pointer
and buffer memories for the procedure presented in
Sect.~\ref{s:sam}\@.

\bparagraph{Building a pointer memory}{p:bui}
The size of the largest block ever sorted by the procedure of
Sect.~\ref{s:sam} will not exceed $m\!=\!n/4$\@.
Using~(\ref{e:pb}) and the fact that the function
$4x/(\log x)^2$ is monotone increasing for $x\!\geq\!8$, we see
that the size of the pointer memory can be bounded by
$P=\flor{4(n/4)/(\log(n/4))^2}=\flor{n/(\log(n/4))^2}$\@. This
will suffice for all sorted blocks.

The pointer memory is built by collecting two contiguous blocks
\pil\ and~\pir\@. The block \pil, placed at the left end of~\ca,
will contain the smallest $P$ elements of the array~\ca, while
\pir, placed at the right end, the largest $P$ elements.

\medbreak
The block \pir\ is created first, by the use of the heapsort with
$t$~root nodes and internal nodes having $t$~sons. The detailed
topology of edges connecting nodes in this kind of heap has been
presented in Sect.~\ref{p:exts}, devoted to extracting sorted
elements at the segment level.

However, there are some substantial differences {}from the
generalized heapsort of Sect.~\ref{p:exts}\@. This time the
branching degree is $t=\ceil{\log n}$\@. Therefore, the heap has
$q\leq 1\!+\!\flor{\log_t n}\leq O(\log n/\log\log n)$ levels.
Here we keep large elements at the root level, instead of small
elements. That is, no node contains an element smaller than any of
its sons. Unlike in Sect.~\ref{p:exts}, no buffer elements are
used here to fill up the holes, the heap structure shrinks in the
standard way, when the largest element is extracted.

The initial building of the heap structure is standard, and agrees
with the heap building in Sect.~\ref{p:exts}\@. It is easy to see
that, for a heap with $n$ elements, branching degree equal
to~$t$, and $q$~levels, the cost of the heap initialization can be
bounded by
$t\!\cdot\!\sum_{i=1}^{q-1}n/t^i < n\!\cdot\!t/(t\!-\!1)\leq
O(n)$
comparisons and
$3\!\cdot\!\sum_{i=1}^{q-1}n/t^i < n\!\cdot\!3/(t\!-\!1)\leq
O(n/\log n)$
moves, using $t\geq\log n$\@.

\medbreak
After building the heap, the routine extracts, $P$~times, the
largest element {}from the heap in the standard way. That is, when
the largest element is extracted, it replaces the element in the
rightmost leaf, which in turn is inserted into the ``proper''
position along the so-called special path, starting {}from the
position of the largest root (just being extracted) and branching
always to the largest son.

The costs of the above routine are straightforward. The trajectory
of the special path can be localized with $q\!\cdot\!(t\!-\!1)$
comparisons, and the new position for the element in the rightmost
leaf can be found by a binary search along this trajectory with
$1\!+\!\flor{\log q}$ comparisons. Summing up, an extraction of
the largest element can be done with
$q\!\cdot\!(t\!-\!1)+(1\!+\!\flor{\log q})$ comparisons, together
with $q\!+\!2$ moves. Using $t\leq O(\log n)$ and
$q\leq O(\log n/\log\log n)$, we get, per a single extraction, at
most $O((\log n)^2/\log\log n)$ comparisons, together with
$O(\log n/\log\log n)$ moves.

If we let the above procedure run till the end, it would sort the
entire array~\ca\ in time $O(n\!\cdot\!(\log n)^2/\log\log n)$\@.
However, the execution is aborted as soon as the largest $P$
elements are collected. Since $P\leq O(n/(\log n)^2)$, the cost of
building the heap becomes dominant, and hence the block~\pir\ is
created with $O(n)$ comparisons and $O(n/\log n)$ moves.

\medbreak
After~\pir, the block~\pil\ is created in the same way, with the
same computational needs of comparisons and moves. Instead of
large elements, here we collect the smallest $P$ elements. In
addition, since \pil\ should be created at the left end of~\ca,
all indices are manipulated in a mirrorlike way, seeing the first
position to the left of~\pir\ as the beginning of the array.

\blemma{l:bui}
Building the pointer memory requires $O(n)$ comparisons and
$O(n/\log n)$ moves.
\elmm

Now the configuration of the array~\ca\ has changed to
$\pil\ca'\pir$, where $\ca'$ denotes the remaining elements, to be
sorted. Before proceeding further, the algorithm verifies, with
a single comparison, whether the largest (rightmost) element in
\pil\ is strictly smaller than the smallest (leftmost) element
in~\pir\@.

If this is not the case, all elements in~$\ca'$ must be equal to
these two elements. Therefore, the algorithm terminates, the
entire array~\ca\ has already been sorted.

Conversely, if \pil\ and~\pir\ pass the test above, they can be
used to imitate a pointer memory consisting of $P$ bits.

\bparagraph{Partition-based sorting}{p:sor}
When the blocks \pil\ and~\pir\ have been created, the zone
$\ca'$ is kept in the form $\ca\s{S}\ca\s{U}$, where $\ca\s{S}$
and~$\ca\s{U}$ represent the sorted and unsorted parts
of~$\ca'\!$, respectively. Each element in~$\ca\s{S}$ is strictly
smaller than the smallest element of~$\ca\s{U}$\@. The routine
described here is a partition-based loop. In the course of the
$i$th iteration, the length of~$\ca\s{U}$ is~$n_i$, with
$n_i\!<\!n_{i-1}$\@. Initially, for $i\!=\!0$, $\ca\s{S}$~is
empty, $\ca\s{U}\!=\!\ca'\!$, and $n_0=n\!-\!2P<n$\@. The loop
proceeds as follows.

\medbreak
First, find~\bs$\!$, an element of rank $\ceil{n_i/4}$
in~$\ca\s{U}$\@. The selection procedure places this element at
the right end of~$\ca\s{U}$, so the configuration of~$\ca'$
changes to $\ca\s{S}\ca'\s{U}\bs\!$\@. Here $\ca'\s{U}$ denotes
a mix of elements in~$\ca\s{U}$, of length $n_i\!-\!1$\@.

Second, $\ca'\s{U}$~is partitioned into two blocks $A\s{<}$
and~$B\s{\geq}$ consisting, respectively, of elements strictly
smaller than~\bs\ and of those greater than or equal
to~\bs$\!$\@. The configuration of the array thus changes to
$\ca\s{S}A\s{<}B\s{\geq}\bs\!$\@. The respective lengths of
$A\s{<}$ and~$B\s{\geq}$ will be denoted here by $n_{i,\st{<}}$
and~$n_{i,\st{\geq}}$\@. Note that, even for a large
block~$\ca\s{U}$, we may obtain a very short block~$A\s{<}$, since
many elements may be equal to~\bs$\!$\@. In fact, the
block~$A\s{<}$ may even be empty, of length
$n_{i,\st{<}}\!=\!0$\@.

Third, sort the block~$A\s{<}$ by the procedure described in
Sect.~\ref{s:sam}, using some initial segments of \pil\
and~\pir\ as a pointer memory and of $B\s{\geq}$ as a buffer
memory, with \bs\ as a buffer separator. This is possible, since
\bs~has been selected as an element of rank $\ceil{n_i/4}$, and
hence $n_{i,\st{<}}\leq\ceil{n_i/4}\!-\!1\leq n_i/4$, with
$n_{i,\st{<}}\!+\!n_{i,\st{\geq}}\!+\!1=n_i$\@. But the required
size of buffer is only
$3n_{i,\st{<}}\!-\!1\leq 3/4\!\cdot\!n_i\!-\!1=
n_i\!-\!1\!-\!n_i/4\leq n_i\!-\!1\!-\!n_{i,\st{<}}=
n_{i,\st{\geq}}$\@.
Therefore, the block~$B\s{\geq}$ of length~$n_{i,\st{\geq}}$ is
sufficiently long. Similarly, the required number of bits for
pointers is
$\flor{4n_{i,\st{<}}/(\log n_{i,\st{<}})^2}\leq
\flor{4(n/4)/(\log(n/4))^2}= P$,
and hence the pointer memory is also sufficiently large. (If
$n_{i,\st{<}}\!\leq\!2^{16}\!$, $A\s{<}$~is sorted as a short
block).

Fourth, restore the sorted order in \pil\ and~\pir, by clearing
all bits of the pointer memory to zero. Among others, this is
required because the procedure of Sect.~\ref{s:sam} will also be
used in subsequent iterations, when it assumes that all bits are
initially cleared.

Fifth, after sorting~$A\s{<}$, the configuration of~$\ca'$ is
$\ca\s{S}A\s{<,S}B'\s{\geq}\bs\!$, where $A\s{<,S}$ denotes the
sorted version of the block~$A\s{<}$ and $B'\s{\geq}$~a~mixed up
version of~$B\s{\geq}$\@. Now put the first element
in~$B'\s{\geq}$ aside and move~\bs\ to the first position
after~$A\s{<,S}$\@. After that, collect all elements smaller than
or equal to~\bs\ to the left part of~$B'\s{\geq}$, processing also
the element put aside. Since $B\s{\geq}$~did not contain elements
strictly smaller than~\bs$\!$, this actually partitions
$B'\s{\geq}$ into two blocks $A\s{=}$ and~$B\s{>}$ consisting,
respectively, of elements equal to~\bs\ and of those strictly
greater than~\bs$\!$, of respective lengths $n_{i,\st{=}}$
and~$n_{i,\st{>}}$\@. Clearly,
$n_{i,\st{=}}\!+\!n_{i,\st{>}}\!=\!n_{i,\st{\geq}}$\@. The
configuration has changed to
$\ca\s{S}A\s{<,S}\bs\!A\s{=}B\s{>}$\@.

Sixth, observe that $\ca\s{S}A\s{<,S}\bs\!A\s{=}$ and $B\s{>}$ can
be viewed as ``new'' variants of blocks $\ca\s{S}$
and~$\ca\s{U}$\@. Thus, we can start a new iteration, with
$B\s{>}$ as a new block~$\ca\s{U}$, of length
$n_{i+1}=n_{i,\st{>}}$\@. The above process is iterated until the
length of unsorted part drops to~$2^{16}\!$, or below. This
residue is then sorted as a short block, without using a buffer or
pointers, which will be described later, in Sect.~\ref{p:sho}\@.

\medbreak
Now we can derive computational costs. First, recall that \bs~has
been selected as an element of rank $\ceil{n_i/4}$, and hence
$n_{i+1}=n_{i,\st{>}}\leq n_i\!-\!\ceil{n_i/4}\leq
3/4\!\cdot\!n_i$\@.
Taking into account that $n_0\!\leq\!n$, we get
$n_i\leq (3/4)^i\!\cdot\!n$, for each $i\!\geq\!0$\@. This gives
that
\bequations{e:it}
  \textstyle\sum_{i=0}^{\ci-1}n_i &\leq& 4n\,,\\
  \ci &\leq& O(\log n)\,,
\eqts
where $\ci$ denotes the number of iterations. Second, it is easy
to see that
\bequation{e:sum}
  \textstyle\sum_{i=0}^{\ci-1}
    (n_{i,\st{<}}\!+\!1\!+\!n_{i,\st{=}}) + n_{\ci}\leq n\,,
\eqtn
since, in different iterations, the final locations occupied by
$A\s{<,S}$, \bs$\!$, and~$A\s{=}$, do not overlap. Here $n_{\ci}$
denotes the length of the residual short block.

Let us now present the costs for the $i$th iteration. Selection
of~\bs$\!$, an element of the given rank in a block of
length~$n_i$, costs $O(n_i)$ comparisons and
$\varepsilon\!\cdot\!n_i$ moves, by~\cite{GK01}\@. Partitioning
of $\ca'\s{U}$ into blocks $A\s{<}$ and~$B\s{\geq}$ can be done
with $n_i$ comparisons and $2n_{i,\st{<}}\!+\!1$ moves, since the
length of~$\ca'\s{U}$ is~$n_i\!-\!1$, and the number of collected
elements, strictly smaller than~\bs$\!$, is~$n_{i,\st{<}}$\@. The
cost of sorting the block~$A\s{<}$ is bounded by
$2n_{i,\st{<}}\!\cdot\!\log n_{i,\st{<}} +
O(n_{i,\st{<}}\!\cdot\!(\log n_{i,\st{<}})^{4/5})\leq
2n_{i,\st{<}}\!\cdot\!\log n +
O(n_{i,\st{<}}\!\cdot\!(\log n)^{4/5})$
comparisons and $(11\!+\!\varepsilon)\!\cdot\!n_{i,\st{<}}$ moves,
by Thm.~\ref{t:sum}\@. Sorting of the block~$A\s{<}$ is followed
by restoring the sorted order in \pil\ and~\pir, by clearing all
bits, which costs $O(P)\leq O(n/(\log n)^2)$ comparisons, together
with the same number of moves. Positioning~\bs\ to the right
of~$A\s{<,S}$ requires only $2$ element moves. Finally, the $i$th
iteration is concluded by partitioning $B'\s{\geq}$ into blocks
$A\s{=}$ and~$B\s{>}$, with at most $n_{i,\st{\geq}}\!\leq\!n_i$
comparisons and $2n_{i,\st{=}}\!+\!1$ moves, since the length
of~$B'\s{\geq}$ is~$n_{i,\st{\geq}}$, and the number of collected
elements, equal to~\bs$\!$, is~$n_{i,\st{=}}$\@. The cost of
sorting the residual short block does not exceed the bounds for
the standard case;
$2n_{\ci}\!\cdot\!\log n_{\ci} + 6.25n_{\ci}\leq
2n_{\ci}\!\cdot\!\log n + O(n_{\ci}\!\cdot\!(\log n)^{4/5})$
comparisons and
$9.75n_{\ci}\leq (11\!+\!\varepsilon)\!\cdot\!n_{\ci}$ moves. (See
Sect.~\ref{p:sho} below).

Now we can sum the above costs over all iterations, using
(\ref{e:it}) and~(\ref{e:sum})\@. For the number of comparisons,
this gives
\bdisplay
  C(n) &\leq& \textstyle\sum_{i=0}^{\ci-1} n_i\!\cdot\!O(1) +
    \textstyle\sum_{i=0}^{\ci-1}
      n_{i,\st{<}}\!\cdot\!(2\log n\!+\!O((\log n)^{4/5})) +
    \textstyle\sum_{i=0}^{\ci-1} O(n/(\log n)^2)\\
  && {}+ n_{\ci}\!\cdot\!(2\log n\!+\!O((\log n)^{4/5}))\\
  &\leq& O(n) +
    (\textstyle\sum_{i=0}^{\ci-1}
      (n_{i,\st{<}}\!+\!1\!+\!n_{i,\st{=}}) + n_{\ci})\cdot
      (2\log n\!+\!O((\log n)^{4/5})) +
    O(n/\log n)\\
  &\leq& O(n) + n\!\cdot\!(2\log n\!+\!O((\log n)^{4/5})) +
    O(n/\log n)\\
  &\leq& 2n\!\cdot\!\log n + O(n\!\cdot\!(\log n)^{4/5})\,.
\edsl
For the number of moves, we get
\bdisplay
  M(n) &\leq&
    \textstyle\sum_{i=0}^{\ci-1} \varepsilon\!\cdot\!n_i +
    \textstyle\sum_{i=0}^{\ci-1}
      (13\!+\!\varepsilon)\!\cdot\!n_{i,\st{<}} +
    \textstyle\sum_{i=0}^{\ci-1} 2n_{i,\st{=}} +
    \textstyle\sum_{i=0}^{\ci-1} O(n/(\log n)^2)\\
  && {}+ (11\!+\!\varepsilon)\!\cdot\!n_{\ci}\\
  &\leq& \varepsilon\!\cdot\!n +
    (\textstyle\sum_{i=0}^{\ci-1}
      (n_{i,\st{<}}\!+\!1\!+\!n_{i,\st{=}}) + n_{\ci})\cdot
      (13\!+\!\varepsilon) +
    O(n/\log n)\\
  &\leq& \varepsilon\!\cdot\!n + n\!\cdot\!(13\!+\!\varepsilon) +
    O(n/\log n)\\
  &\leq& (13\!+\!\varepsilon)\!\cdot\!n\,,
\edsl
where $\varepsilon\!>\!0$ is an arbitrarily small, but fixed, real
constant. The above analysis did not include the costs of the
initial building of pointer memory. However, by Lem.~\ref{l:bui},
this can be done with only $O(n)$ comparisons and $O(n/\log n)$
moves, and hence the bounds displayed above represent the total
computational costs of the algorithm.

\btheorem{t:sor}
The given array, consisting of $n$ elements, can be sorted
in-place by performing at most
$2n\!\cdot\!\log n + o(n\!\cdot\!\log n)$ com\-par\-i\-sons and
$(13\!+\!\varepsilon)\!\cdot\!n$ element moves, where
$\varepsilon\!>\!0$ denotes an arbitrarily small, but fixed, real
constant. The number of auxiliary arithmetic operations with
indices is bounded by $O(n\!\cdot\!\log n)$\@.
\ethm

\bparagraph{Handling short blocks}{p:sho}
The algorithm presented above needs a procedure capable of sorting
blocks of small lengths, namely, with
$m\!\leq\!2^{16}\!=\!65536$\@. This is required, among others, to
sort blocks~$A\s{<}$ that are short. We could sweep the problem
under the rug by saying that ``short'' blocks can, ``somehow,'' be
sorted with $O(1)$ comparisons and moves, since they are of
constant lengths. However, the upper bounds presented by
Thm.~\ref{t:sum} in Sect.~\ref{p:sum} require some more details,
especially for $(11\!+\!\varepsilon)\!\cdot\!m$, the number of
moves. Last but not least, these lengths are important in
practice.

One of the possible simple solutions is to use our version of
heapsort, with $5$~roots and internal nodes having $5$~sons. Using
the analysis presented in Sect.~\ref{p:bui}, devoted to building
a pointer memory, for $t\!=\!5$, $m\!\leq\!2^{16}\!$, and hence
for at most $q\leq 1\!+\!\flor{\log_t m}\leq 7$ levels, one can
easily verify that we shall never use more than
$2m\!\cdot\!\log m + 6.25m$ comparisons or $9.75m$ moves. (These
bounds are not tight, we leave further improvement to the reader).

\bparagraph{An alternative solution}{p:alt}
As pointed out at the end of Sect.~\ref{p:exts}, devoted to
extracting sorted elements {}from segments, we could use a heap
structure with four levels, instead of five, in a segment. This
slightly reduces the number of moves, but increases the number of
comparisons. The detailed argument parallels the proof of
Thm.~\ref{t:sor}, and hence it is left to the reader.

\bcorollary{c:sor}
The given array, consisting of $n$ elements, can be sorted
in-place by performing at most
$6n\!\cdot\!\log n + o(n\!\cdot\!\log n)$ com\-par\-i\-sons and
$(12\!+\!\varepsilon)\!\cdot\!n$ element moves, where
$\varepsilon\!>\!0$ denotes an arbitrarily small, but fixed, real
constant.
\ecrr

\bsection{Concluding Remarks}{s:c}
We have described the first in-place sorting algorithm performing
$O(n\!\cdot\!\log n)$ comparisons and $O(n)$ element moves in the
worst case, which closes a long-standing open problem.

However, the algorithms presented in Thm.~\ref{t:sor} and
Cor.~\ref{c:sor} do not sort stably, since the order of buffer
elements may change. If some elements used in buffers are equal,
their original order cannot be recovered. This leaves us with
a fascinating question:
\begin{quotation}\em\noindent
  Does there exist an algorithm operating in-place and performing,
  in the worst case, at most $O(n\!\cdot\!\log n)$ comparisons,
  $O(n)$ moves, $O(n\!\cdot\!\log n)$ arithmetic operations, and,
  at the same time, sorting elements stably, so that the relative
  order of equal elements is preserved?
\end{quotation}

At the present time, we dare not formulate any conjectures about
this problem. The best known algorithm for stable in-place sorting
with $O(n)$ moves is still the one presented in~\cite{MR96s},
performing $O(n^{1+\varepsilon})$ comparisons in the worst case.

\medbreak
We are also firmly convinced that the upper bounds of
Thm.~\ref{t:sor} and Cor.~\ref{c:sor} are not optimal and can be
improved, which is left as another open problem.

\bibliography{sorting}%
\bibliographystyle{abbrv}
\end{document}